\documentclass[%
 reprint,
 amsmath,amssymb,
 aps,
floatfix,
]{revtex4-1}

\usepackage{graphicx}
\usepackage{dcolumn}
\usepackage{bm}

\begin{document}

\preprint{APS/123-QED}

\title{Selectively strong coupling MoS$_2$ excitons to a metamaterial at room temperature}

\author{Harshavardhan R Kalluru}
\email{kallurureddy@iisc.ac.in}
 \affiliation{Department of Physics, Indian Institute of Science, Bangalore 560012, India}
\author{Jaydeep K Basu}
 \affiliation{ Department of Physics, Indian Institute of Science, Bangalore 560012, India}
\date{\today}

\begin{abstract}
Light emitters in vicinity of a hyperbolic metamaterial (HMM) show a range of quantum optical phenomena from spontaneous decay rate enhancement to strong coupling. In this study, we integrate monolayer molybdenum disulfide (MoS$_2$) emitter in near field region of HMM. The MoS$_2$ monolayer has A and B excitons, which emit in the red region of visible spectrum. We find that the B excitons couple to HMM differently compared to A excitons. The fabricated HMM transforms to a hyperbolic dispersive medium at 2.14 eV, from an elliptical dispersive medium. The selective coupling of B excitons to the HMM modes is attributed to the inbuilt field gradient of the transition. The B exciton energy lies close to the transition point of the HMM, relative to A exciton. So, the HMM modes couple more to the B excitons and the metamaterial functions as selective coupler. The coupling strength calculations show that coupling is 2.5 times stronger for B excitons relative to A excitons. High near field of HMM, large magnitude and the in-plane transition dipole moment of MoS$_2$ excitons, result in strong coupling of B excitons and formation of hybrid light{-}matter states. The measured differential reflection and photoluminescence spectra indicate the presence of hybrid light-matter states i.e. exciton-polaritons. Rabi splitting of 143.5 meV$\pm$14.4 meV at room temperature is observed. The low temperature photoluminescence measurement shows mode anticrossing, which is characteristic feature of hybrid states. Our results show that the HMM works as a energy selective coupler for multi-excitonic systems as MoS$_2$.
\end{abstract}

\maketitle


\section{INTRODUCTION}

The hyperbolic metamaterial (HMM) undergoes an optical topological transition (OTT) and the iso-frequency surface of HMM changes from an ellipsoid to a hyperboloid at the transition point. The volume under the iso-frequency surface is the photonic density of states (PDoS). The hyperbolic iso-frequency surface is non-integrable. As a consequence of OTT, the PDoS of HMM diverges.\cite{jacob_kim_naik_boltasseva_narimanov_shalaev_2010} Due to diverging PDoS and nature of propagating hyperbolic modes\cite{Yao2008}, the HMM system has been studied extensively for controlling light-matter interaction. 

The evanescent modes with large wave-vector propagating through the volume of HMM enable imaging of super-resolution imaging of sub-diffraction size objects in far field.\cite{Liu2007, sridhar2010} HMM has been used to sense pico molar concentrations of bio molecules.\cite{Kabashin2009, KVSreekanth} The interface between HMM and a photonic crystal as a Bragg mirror, can support optical Tamm states.\cite{Bikbaev:20} The optical Tamm state wave vectors lie with in the light cone and can be directly excited by incident light beam.\cite{Shelykh2007} This has applications in broadband absorbing materials, optical filters\cite{WU2021110680} and photo-voltaic devices.\cite{Wu:21}

It has been demonstrated that a collection of emitters on hyperbolic materials show strong coupling near the OTT set-in point.\cite{Biehs_2018} Modification of spontaneous emission rate of dye molecules,\cite{DianeJRoth2017} rare earth ions\cite{LopezMorales2021} and strong coupling of semiconductor quantum dots\cite{Chaitanya2017} in vicinity of HMM is reported. The transition energy can be tuned by changing the fabrication parameters.

Monolayer molybdenum disulphide (MoS$_2$) is the light emitter used for coupling with the HMM. MoS$_2$ belongs to a class of layered Van der Waals materials called as transition{-}metal dichalcogenides (TMDCs).These materials can be exfoliated mechanically to the monolayer limit. The monolayer MoS$_2$ is a direct band-gap semi conductor.\cite{Mak2010} It has applications in light emitting devices\cite{Sundaram2013}, optical valley control devices\cite{Mak2012}, energy storage systems\cite{Jiang2015} and piezo-electric\cite{Wu2014} devices. Monolayer MoS$_2$ system has neutral excitons, charged excitons, localized excitons and dark excitons.\cite{Mueller2018}

The orientation of transition dipole moment (TDM) of emitters relative to the metal{-}dielectric interface of HMM determines the PDoS experienced by the emitters\cite{Shalaginov2015} and the efficiency of coupling.\cite{Kala2020} If the TDM is oriented normal to metal-dielectric interface of HMM, the coupling is optimum to the plasmonic modes. 

Due to the in-plane orientation\cite{Schuller2013} of TDM, the monolayer MoS$_2$ excitons are well suited for coupling to silver nanowire based HMM. The maximum field of the TDM is normal to the plane of monolayer. So the PL emission of MoS$_2$ monolayer is ideal for driving the longitudinal plasma oscillations in silver nanowires of HMM. The high TDM magnitude\cite{Wang2019} of TMDC excitons $\sim$ 50 debye is also an additive factor for optimal coupling.

Strong coupling\cite{Khitrova2006} of emitters and HMM is observed, when the coupling strength dominates the emitter spontaneous decay rate and HMM losses. Strong coupling results in Rabi splitting in the frequency domain. The magnitude and orientation of the emitter TDM ($\mu$) determines the energy decay rate into vacuum and HMM. The HMM PDoS determines the field strength (E). The PDoS is highest near OTT set-in point and gradually decreases as the exciton emission peak moves away from the transition point. So the both factors determine the coupling coefficient $g$ of a single emitter, given by\cite{D.G.Baranov2018}
\[ g=\frac{\mu E}{\hbar} \tag{1} \]

We show that the MoS$_2$ B excitons are strongly coupled to HMM and the Rabi Splitting is 143.5 meV $\pm$ 14.4 meV at room temperature. Similar value of Rabi splitting (101 meV) is reported for WS$_2$ coupled metallic nano structures.\cite{SWang2016}. Rabi splitting of 58 meV at 77 K is reported for the MoS$_2$ monolayer coupled plasmonic lattice system.\cite{WLiu2016} The lithographic processes used for fabricating nano scale plasmonic systems are expensive. Silver nanowire based HMM, used in this study is made through standard electrochemical processes of aluminum metal finishing industry\cite{Sheasby1987} is an inexpensive alternative. The fabricated HMM is an industrially scalable coupler, which can selectively strong couple MoS$_2$ excitons at room temperature.

\section{EXPERIMENTAL METHODS}

The HMM is fabricated on a glass substrate for room temperature measurements. Exfoliated MoS$_2$ monolayer is transferred on to the HMM with 10 nm polymer spacer. The Raman and white light reflection spectra are collected with a confocal microscope (Witec alpha 300). A 532 nm laser diode is used for Raman spectra measurement and the spectra are collected with 100x magnification and 0.9 numerical aperture (NA) bright field objective. The signal from sample is relayed to a grating coupled Peltier cooled CCD. For low temperature photoluminescence measurement the sample was fabricated on a Silicon (with 300 nm oxide) substrate. The sample is placed in a closed cycle cryostat (Montana), which is capable to reach till 6 K. The cryostat is placed directly under a microscope system (Horiba labram) for PL measurement in reflection mode. The PL spectra are excited with a 532 nm frequency doubled Nd:YAG laser and are collected with a 50x magnification and 0.5 NA objective and relayed on to a grating coupled Peltier cooled CCD.
\section{RESULTS AND DISCUSSION}
\subsection{HYPERBOLIC METAMATERIAL}
The HMM is fabricated as per an earlier work from our group\cite{Yadav2020}, by growing silver nanowires in a porous aluminum oxide film, through electrochemical deposition. (supporting material S1) The HMM is characterised by scanning electron microscopy (SEM) and atomic force microscopy (AFM), as shown in Fig. 1. The metamaterial undergoes OTT and transforms from a lossy dielectric medium to hyperbolic dispersive medium. The transition shown in Fig. 2(a), is calculated according to effective medium theory (EMT) approximation\cite{Shekhar2014} of HMM, with silver metal filling fraction 0.15. The absorption coefficient of a system is directly proportional to differential reflectivity (supporting material S4). The  unpolarised white light differential reflection (dR) spectra of HMM  are measured and shown in Fig. 2(b). The OTT of HMM is seen as a smooth transition\cite{Krishnamoorthy2012}, which is attributed to the losses in HMM.

\begin{figure}[htp]
\includegraphics{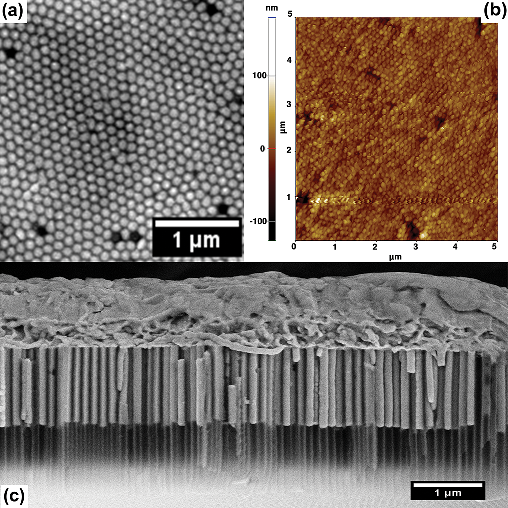}
  \caption{(a) shows the SEM image of region etched with focused ion beam (FIB). (b) shows an AFM image of 5 {\textmu}mx5 {\textmu}m region of the roughened side of HMM with 10 nm spacer. (c) shows the cross sectional SEM image of silver nanowires grown in Aluminum oxide template.}
  \label{fgr:1}
\end{figure}

The HMM dR spectra has a broad feature between 1.9 eV and 2.1 eV. To find out the critical point, where transition occurs, the first order derivative of measured dR is shown in the inset of Fig. 2(b). The critical point is at 2.1003 eV $\pm$ 0.8 meV, which is in good agreement with the calculated EMT transition point 2.1378 eV. So the EMT description of HMM is accurate. From a theoretical perspective, the light free space wavelength (in this paper $\lambda_o>$ 500nm) is much larger than the period (133 nm) of the unit cell. So the HMM should behave as an effective medium, which is what we see in dR spectra.  The monolayer MoS$_2$ has three luminescent excitons\cite{Steven2013, Marcos2015} namely, A (about 1.8 eV),B (at 2.0 eV) and C (at 2.8 eV). The C exciton is not useful for coupling, as it is in elliptical dispersive region of HMM.
\subsection{COUPLING STRENGTH CALCULATIONS}
The selective strong coupling of B excitons to HMM can be understood by considering the difference in PDoS at the exciton positions and the relative oscillator strength of B and A excitons. The PDoS parameter (P) of a lossy hyperbolic half space\cite{Jacob2012} is calculated from reported values of refractive indices and extinction coefficients of Silver and Alumina.\cite{Johnson1972}\[ \textup{P}= \frac{2\sqrt{\epsilon_{xx}|\epsilon_{zz}|}}{(1+\epsilon_{xx}|\epsilon_{zz}|)} + \epsilon^{\prime\prime}\frac{(\epsilon_{xx}-|\epsilon_{zz}|)}{(1+\epsilon_{xx}|\epsilon_{zz}|)^2} \tag{2}\]The measured absorption data is incorporated in the PDoS calculation as the loss factor ($\epsilon^{\prime\prime}$). Here Z axis is the growth direction of nanowires and $\epsilon_{xx}$ , $\epsilon_{zz}$ are effective permittivity values along X and Z axes. The calculated PDoS is shown in Fig. 2(d). 
The PDoS is directly proportional to the electric field (E).\cite{Li_2021} So the ratio of HMM electric field at B (2.0222 eV) to A (1.8463 eV) exciton position is given by \[\frac{\textup{P}_\textup{B}}{\textup{P}_\textup{A}}=\frac{\textup{E}_\textup{B}}{\textup{E}_\textup{A}} =\text{2.18} \tag{3} \]
The oscillator strength ($\mu^2$) of an exciton\cite{Hilborn1982} is directly proportional to the absorption coefficient ($\alpha$). The relative oscillator strength of B to A excitons is extracted from measured monolayer MoS$_2$ dR spectra.\[\frac{\mu^2_\textup{B}}{\mu^2_\textup{A}} = \frac{\alpha_\textup{B}}{\alpha_\textup{A}} =\text{1.49} \tag{4} \]
\begin{figure}[htp]
\includegraphics{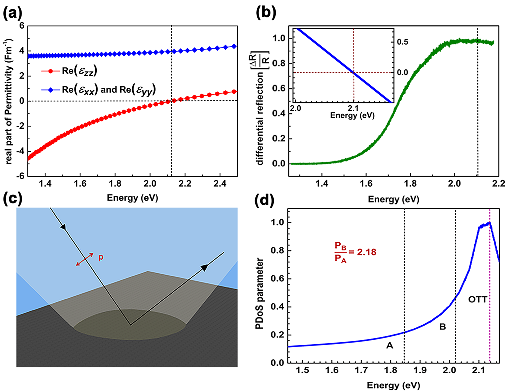}
  \caption{(a) shows the calculated OTT of HMM,(b) shows the measured dR spectra of HMM and the inlay shows the first order derivative of dR spectra (c) shows the schematic for HMM white light reflection (d) shows the calculated PDoS parameter (P) for HMM. The lines indicate the positions of A and B excitons}
  \label{fgr:2}
\end{figure}
The exciton population ($N$) is directly proportional to the PL intensity of MoS$_2$ excitons at room temperature. (supporting information S5) The ratio turns out as \[ \frac{N_\textup{B}}{N_\textup{A}}=\frac{I_\textup{B}}{I_\textup{A}}=\text{0.88}\tag{5} \]The coupling coefficient ($g$) for a population of $N$ excitons\cite{Torma2014,Jussi2018}  is given by $\hbar{g}=\sqrt{N}\mu\textup{E}$, where $\mu$ is the excitonic transition dipole moment and E is the HMM electric field. The ratio of coupling coefficients is calculated by combining equations (3),(4) and (5).
\[ \frac{g_\textup{B}}{g_\textup{A}} = \sqrt{\frac{N_\textup{B}}{N_\textup{A}}} \frac{\mu_\textup{B}\textup{P}_\textup{B}}{\mu_\textup{A}\textup{P}_\textup{A}}=\text{2.50} \tag{6} \]The coupling at B exciton is 2.5 times larger than the coupling at A exciton. So, the HMM can act as a selective coupler for MoS$_2$ excitons.
\subsection{COUPLED MoS$_2${-}HMM SYSTEM}
Bulk MoS$_2$ pieces are mechanically exfoliated using scotch tape until monolayer limit is reached. The monolayer is transferred on to HMM by dry visco-elastic gel stamping.\cite{Sarkar2019,PradeepaHL2020} (supporting material S2 and S3)

The measured dR spectra for HMM, monolayer MoS$_2$ system on glass and monolayer MoS$_2$ on HMM are plotted in Fig. 3(b). The dR spectra of MoS$_2$ monolayer on glass has two peaks at  1.8463 eV $\pm$ 0.8 meV  and  2.0222 eV $\pm$ 0.8 meV, corresponding to A and B excitons respectively. An extra peak at 1.9280 eV $\pm$ 0.8 meV is observed between the positions of A and B excitons, in dR spectra of coupled monolayer MoS$_2$-HMM system.

The measured PL spectra for monolayer MoS$_2$ on HMM and monolayer MoS$_2$ system on Silicon with varying temperature are shown in Fig. 4(a) and Fig. 4(b), respectively. The room temperature (289 K) PL spectral envelope of monolayer MoS$_2$ on HMM is modified and an extra peak ($\omega_{-}$) appears at 1.8955 eV $\pm$ 7.2 meV. So a new peak is observed in both dR (absorption) and PL (emission) spectra, which confirms strong coupling.\cite{Leng2018}. Only three spectra at different temperatures are shown for brevity. (supporting material S6)

MoS$_2$ monolayer Debye temperature\cite{Peng2016} is 262 K which is close to room temperature. So the PL peak position is sensitive to cooling, as PL is a band-edge emission process. The MoS$_2$ A and B exciton PL peak positions blue shift on cooling to cryogenic temperatures,\cite{Korn2011} where as the plasmonic modes are temperature insensitive. Lowering temperature can detune\cite{Reithmaier2004} the PL spectra and is used to visualize anticrossing\cite{Y.Chen2017} behaviour of the MoS$_2$-HMM coupled system.

The losses are always significant in a plasmonic system as HMM due to the scattering\cite{Khurgin2015} of electrons by phonons, electron-electron scattering and direct scattering by surface plasmon polaritons (SPPs). When cooled down to cryogenic temperatures, phonon-electron scattering reduces drastically, which results in longer\cite{Hummer2013} propagation lengths of SPPs. Essentially a fraction of HMM losses can be minimized, by cooling the system, which is an added advantage of low temperature PL measurement.
\begin{figure}[htp]
\includegraphics{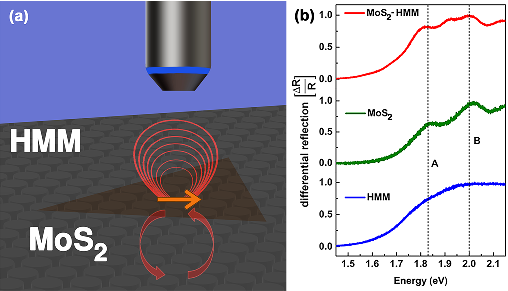}
  \caption{(a) shows the configuration for coupling MoS$_2$ excitonic transition dipole moment to HMM (b) shows the dR spectra of HMM, monolayer MoS$_2$ and coupled monolayer MoS$_2${-}HMM system respectively}
  \label{fgr:3}
\end{figure}
So, temperature dependent PL detuning measurement can help to understand whether A or B exciton is strongly coupled to HMM. Mode anticrossing is observed for strongly coupled polariton modes \cite{Wersall2017, Dovzhenko2018}. While detuning, one polariton mode moves towards the exciton line and the other polariton mode moves away from the exciton line.\cite{Novotny2010, Kleemann2017}

We focus on PL features in the region from 1.8 eV to 2.1 eV in this paper, as the features related to A and B excitons appear in it. Monolayer MoS$_2$ PL peak position is strongly dependent\cite{YSun2017} on substrate and dielectric environment. The refractive index of HMM also changes with temperature. To make sense of the low temperature PL data, two control measurements (supporting material S7) are done, to account for the temperature dependent refractive index of environment. Rhodamine B dye molecule PL peak position is independent\cite{Hogg2018} of temperature. So for control measurements, Rhodamine B dye solution was drop casted on to both HMM and Silicon with spacer. Monolayer MoS$_2$ was transferred on to Silicon with spacer as a reference sample for exciton peak blue shift. 

The collected PL spectra, of all the samples are fitted with lorentzian function and the corresponding peak positions are obtained.The MoS$_2$ PL peak shifts observed on HMM and Silicon sample as are corrected with respective control samples with Rhodamine B dye. The measured temperature points range from 6 K $\pm$ 0.05 K to room temperature. The B exciton peaks for MoS$_2$ on Silicon are fitted with a line. The corrected peak positions are shown in Fig. 4(c) and Fig. 4(d).
\begin{figure}[htp]
\includegraphics{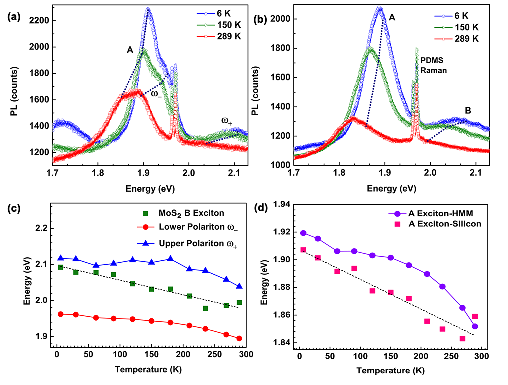}
  \caption{(a) shows the measured temperature dependent PL of monolayer MoS$_2$ coupled to HMM. (b) shows the  temperature dependent PL of monolayer MoS$_2$ on Silicon substrate.(c) and (d) show the evolution of PL peak features of both monolayer MoS$_2$ and monolayer MoS$_2$ coupled to HMM with lowering temperature.}
  \label{fgr:4}
\end{figure}
To visualise the evolution of peaks in Fig. 4(c), the energies of the peaks adjacent to the B exciton are subtracted with the MoS$_2$ B exciton energy and are shown in Fig. 5. With lowering temperature, the subtracted upper polariton branch [$\omega_\textup{+}{-}\omega_{\textup{B}}$] moves towards the B exciton line and the subtracted lower polariton branch [$\omega_{-}{-}\omega_\textup{B}$] moves away from the B exciton line.
\begin{figure}[htp]
\includegraphics{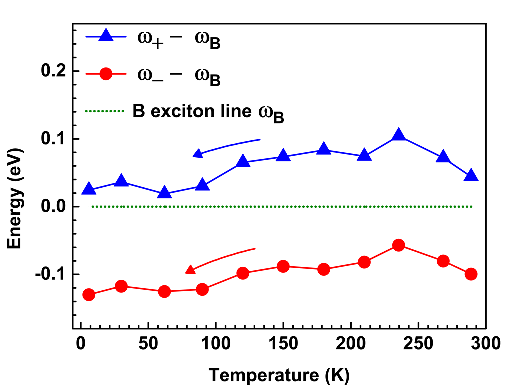}
  \caption{shows the evolution of peaks with respect to B exciton line with lowering temperature}
  \label{fgr:5}
\end{figure}

The anticrossing of modes, confirms that the B excitons of monolayer MoS$_2$ are strongly coupled with HMM plasmonic modes and form exciton-plasmon-polaritons. The A excitons peaks on HMM and on control Silicon lie with in the error bar of 14.4 meV. (supporting material S9) Anticrossing is not observed for A excitons. So they are not strongly coupled with HMM modes. The B excitons are selectively strong coupled to the HMM modes. The separation between the upper and lower polariton branches is considered as Rabi splitting. From Fig. 4(c), its value turns out as 143.5 meV$\pm$14.4 meV at room temperature.

We have shown due to the inbuilt field gradient of metamaterial, the nature of coupling is different for B and A excitons. The B excitons are strongly coupled to metamaterial at room temperature. Apart from the coupling strength considerations, the monolayer MoS$_2$ is a direct band gap semi conductor. So, the polaritons can be encoded directly by a laser beam. The coupled MoS$_2$-HMM system has application as an encoder in optical information systems.

\section{CONCLUSION}
The coupled MoS$_2$-HMM system is characterized with differential reflection and temperature dependent photoluminescence measurements. The mode anticrossing confirms that the B excitons of monolayer MoS$_2$ are strongly coupled to HMM. At room temperature, the measured Rabi splitting is 143.5 meV$\pm$14.4 meV. We show that the metamaterial functions as an energy selective coupler for MoS$_2$ A and B excitons. The relative coupling strength is 2.50 for B to A excitons. 

\section{Supporting material}
\subsection{SAMPLE FABRICATION PROCEDURE}
The Silver nanowire based hyperbolic metamaterial (HMM) is fabricated by two step anodization of 6N purity aluminium metal sheets procured from Alfa Aesar. The aluminium sheet is sonicated in ethanol for 2 minutes to remove any organic residues on surface. The sonicated sheets are washed thrice in 18.2 M$\Omega$ deionised water and dried. The dried sheets are electro polished in 1:4 volumetric mixture of ethanol and perchloric acid at 25 V and 3 A current for 2 minutes. 
A thin line of high performance acrylic electrolube (HPA) is coated at the interface of polished and unpolished aluminium metal to avoid surface current flow from unpolished to polished region. Along with a graphite cathode, the polished metal sheet is dipped in 0.15M ice cold oxalic solution for Anodization process. The metal sheet is anodized for 14 hours at 50V and 3A for oxide growth. The grown oxide is then etched using a dilute Chromic acid solution (2\% w/V) for 8 hours. Then the metal sheet is cleaned thrice in de-ionised (DI) water. 
The metal sheet is further anodized for 2 hours at 50V and 3A for growing Aluminium oxide. The second anodized metal sheet is placed in a Teflon jacket and the top part of Aluminium oxide is etched using 20 mg/ml sodium hydroxide (NaOH) solution for 15 minutes. The metal is etched using a 1:2 mixture of 0.25M copper chloride (CuCl2) and 0.1M hydrochloric acid (HCl). The final left over aluminium oxide at bottom face is cleaned thrice in DI water. The barrier layer is removed by etching in 6\% v/V Phosphoric acid solution (MERCK product: 345245) for 40 minutes. Then the porous aluminium oxide template is cleaned thrice with DI water and dried. 
A thin gold film is sputtered on to the template to act as an electrode for electro deposition. Silver nanowires are grown in the aluminium oxide template using a deposition solution of 0.1M silver bromide (AgBr), sodium hyposulphite (Na2SO3), Sodium Sulphite (NaSO3). The grown metamaterial is characterised by atomic force microscopy (AFM) and scanning electron microscopy (SEM). 
\begin{figure}[htp]
\includegraphics{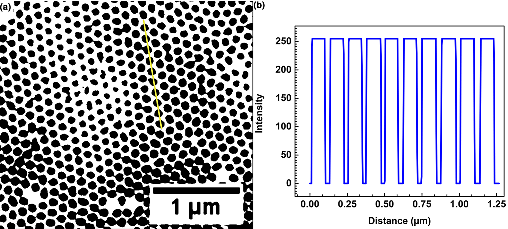}
  \caption{(a) shows the processed 8 bit grayscale image (intensity values 0 to 255) of the raw SEM image from Fig. 2(a) of main part of the manuscript. The yellow colored line is used to extract the line profile of SEM image.  (b) shows the extracted line profile, with the wires saturated at 255.}
  \label{fgr:S1}
\end{figure}
The distance between edges of a saturated segment (Intensity=255) in Fig. S1 (b) is considered as silver nanowire diameter ($d$) and the distance between two consecutive segment centres is considered as inter wire separation ($s$). The measured diameter of the nanowire ($d$) is 54 nm and the wire to wire separation ($s$) is 133 nm. The filling fraction ($f$) considering hexagonal unit cell of HMM is calculated as \[f=\frac{\pi}{2\sqrt{3}}\left(\frac{d}{s}\right)^2 \tag{7}\]
The filling fraction turns out as 0.1495. The unfilled aluminium oxide is removed using a dilute NaOH (0.5 mg/ml) solution. This makes the top surface of HMM roughened. Roughened HMM absorbs light better than a smoothened metamaterial.\cite{Narimanov:13} The top surface roughness of metamaterial is about 250 nm. The bottom surface is plated with 1 micron silver metal due to overgrowth. The plated silver acts as an electrical contact. This ensures the silver nanowire plasmon polariton modes are symmetric\cite{Podolskiy_2005} which is essential for collective response of HMM system. Due to these reasons, the roughened side of metamaterial is chosen for coupling to MoS2 emitter.
Bulk MoS2 crystal from SPI supplies is mechanically exfoliated till monolayer MoS2 is obtained. The monolayer flake is transferred to a commercially available (Gel-Pak) PolyDiMethylSiloxane (PDMS) gel stamp.\cite{Castellanos_Gomez_2014} The PDMS stamp with flake is transferred on to the HMM with 10nm Polystyrene polymer spacer using a micromanipulator. The PDMS stamp is retained on HMM to create an isotropic dielectric environment. For PDMS layer and glass substrate, the refractive index mismatch is less than 0.1. This amplifies the reflection contrast, which enables easy detection of absorption features. 
\subsection{OPTICAL IMAGES AND PL SPATIAL MAP }
\begin{figure}[htp]
\includegraphics{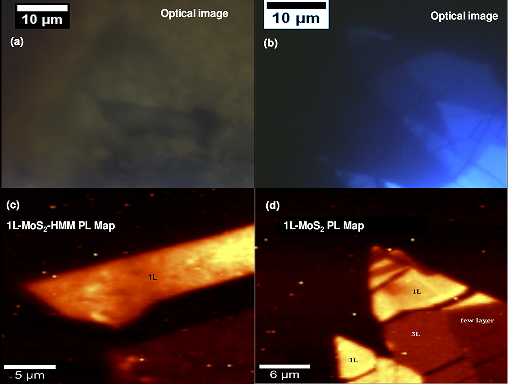}
  \caption{(a) and (b) show the optical images of MoS$_2$ monolayer on HMM system and the control sample MoS$_2$ monolayer on glass and respectively. (c) and (d) show the respective PL spatial maps.}
  \label{fgr:S2}
\end{figure}
\subsection{RAMAN SPECTRUM}
\begin{figure}[htp]
\includegraphics{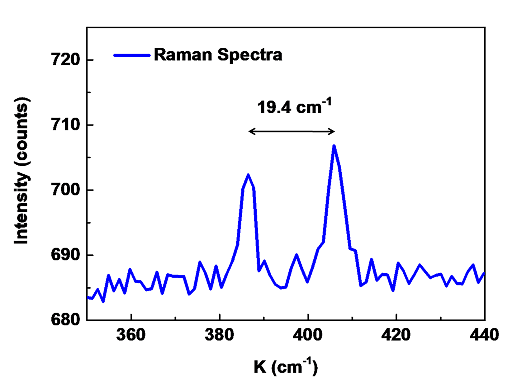}
  \caption{shows the Raman spectra of the MoS$_2$ monolayer on HMM.}
  \label{fgr:S3}
\end{figure}
\subsection{DIFFERENTIAL REFLECTION MEASUREMENT}
The sample absorption can be measured through the white light differential reflection measurement.\cite{MCINTYRE1971417,PhysRevB.90.205422,C4NR03703K} Differential reflection ($\frac{\Delta R}{R}$ ) is measured from the equation (2), where $n_1$ and $n_3$ are refractive indices of media above and below the sample. The absorption coefficient of sample is given by $\alpha_2$ and $n_2$ is refractive index of sample. The sample thickness is $d$.
\[ \frac{\Delta R}{R} = \frac{4dn_1n_2}{n_3^2-n_1^2} \alpha_2 \tag{8} \]
\begin{figure}[htp]
\includegraphics{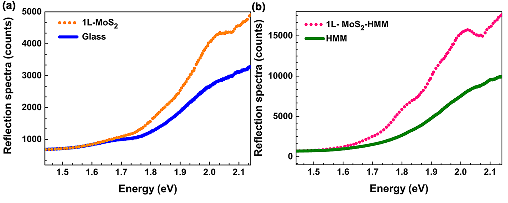}
  \caption{(a) shows the white light reflection spectra of MoS$_2$ monolayer system and glass substrate. (b) shows the white light reflection spectra of MoS$_2$ monolayer - HMM system}
  \label{fgr:S4}
\end{figure}
\subsection{COUPLING COEFFICIENT CALCULATION}
The intensity ($I$) i.e. counts of photoluminescence (PL) spectra is directly proportional to the population of excitons ($N$). The ratio of counts at B exciton to A exciton position is extracted from the fit to PL spectra in Fig. S5 (a). The ratio turns out as 0.88. Similarly from PDoS curve of HMM in Fig. S5 (b), the ratio of density of states is extracted as 2.18.
\begin{figure}[htp]
\includegraphics{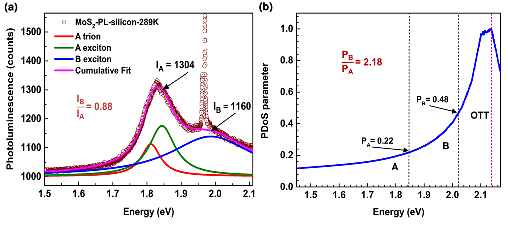}
  \caption{a) shows the fitting of photoluminescence spectra of MoS$_2$ monolayer at room temperature. (b) shows the calculated PDoS of HMM at A and B exciton positions.}
  \label{fgr:S5}
\end{figure}
\subsection{LOW TEMPERATURE PL MEASUREMENT}
\begin{figure}[htp]
\includegraphics{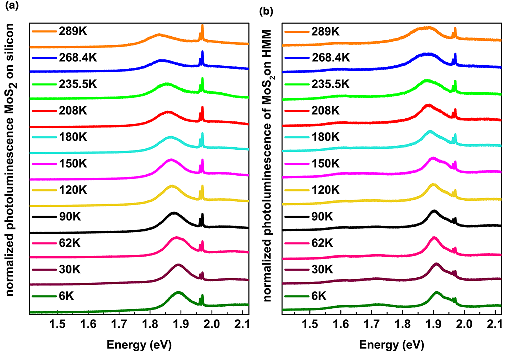}
  \caption{The temperature dependence of the PL spectra of MoS$_2$ monolayer on silicon and MoS$_2$ monolayer on HMM system is shown in (a) and (b) respectively.}
  \label{fgr:S6}
\end{figure}
\subsection{LOW TEMPERATURE PL CALIBRATION}
Rhodamine B(RhB) dye is a strongly luminescent xanthene dye, which exists in its monomer\cite{LOPEZARBELOA1982556} form at micro molar concentration. A dilute solution of 5 µM RhB solution in ethanol is prepared and 10µL of the solution is drop casted on to HMM with Polystyrene(PS) spacer and Silicon with PS spacer. The samples are left to dry in ambient conditions. Low temperature PL measurements are done for both samples. The low temperature PL peak positions of RhB on HMM and Silicon serve for correcting dielectric environment fluctuations with respect to room temperature. The corresponding peak corrections for HMM and Silicon are added to the peaks obtained from the fitted data of MoS$_2$ on HMM and MoS$_2$ on Silicon.
\begin{figure}[htp]
\includegraphics{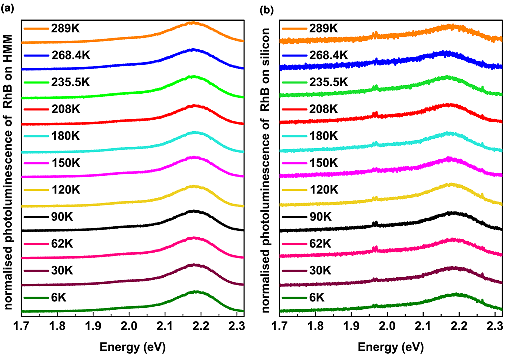}
  \caption{The temperature dependence of the PL spectra of Rhodamine B dye on HMM and Rhodamine B dye on Silicon is shown (a) and (b) respectively.}
  \label{fgr:S7}
\end{figure}
\subsection{TEMPERATURE DEPENDENT PL SPECTRA FITTING}
The PL temperature dependent PL spectra are fitted with Lorentzian function and the peak position information is extracted. The fitted PL spectra of the MoS$_2$ monolayer on Silicon and MoS$_2$ monolayer HMM at 6 K are shown in Fig. S8 (a) and Fig. S8 (b) respectively. The peaks are corrected with Rhodamine PL spectra are shown in Fig. S8 (c) and Fig. S8 (d) respectively. 
Defect bound excitons show up about $\simeq$ 0.2 eV, lower than the A exciton peak position.\cite{Tongay2013} The A and B excitons blue shift with lowering temperatures. The defect bound excitons (D, D1 and D2) evolve as per the previously reported\cite{Saigal2016} trend. The peaks D1 and D2 are due to the defects bound to single and double sulphur vacancies of MoS$_2$.
\begin{figure}[htp]
\includegraphics{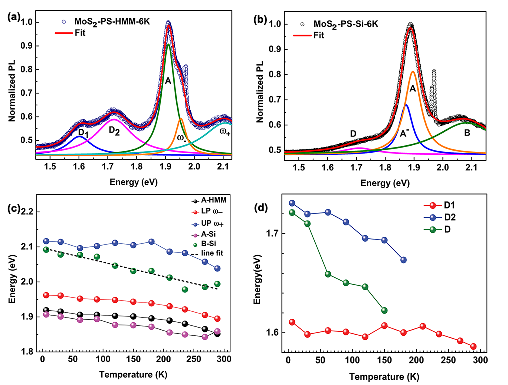}
  \caption{(a) and (b) show the fitting of the evolution of PL spectra of MoS$_2$ monolayer on HMM and MoS$_2$ monolayer on Silicon (Si), respectively. (c) and (d) show the evolution of PL peaks of $\omega_{-}, \omega_{+}$, B, A, defect bound excitons of MoS$_2$ monolayer(D1,D2) on HMM with reference to defect bound excitons(D) on Silicon.}
  \label{fgr:S8}
\end{figure}

\subsection{ERROR ANALYSIS}
The maximum fitting error of PL spectra peaks is 6.8 meV. The grating coupled CCD detector spectral resolution is 0.41 meV. So the error bar for PL peak position is 7.21 meV at worst case scenario. Since differential reflection involves the ratio of two measured spectra with CCD, the error bar for differential reflectivity spectra is 0.82 meV. The temperature dependent PL spectra peak positions are corrected with Rhodamine spectra, so the error bar for Rabi splitting is 14.42 meV.
\begin{acknowledgments}
The authors thank the department of science and technology (DST), India - Nanomission and the funds for improvement of  science and technology (FIST) programmes for financial support. HR Kalluru thanks micro and nano characterization facility (MNCF-CeNSE); advanced facility for microscopy and microanalysis (AFMM), Indian Institute of Science, Bangalore, for access to SEM and FIB facilities. Authors thank Prof. V.M.Menon, City College of New York for discussions on anticrossing measurements.
\end{acknowledgments}
\bibliographystyle{apsrev4-1}
\bibliography{apssamp.bib}
\end{document}